\documentclass[preprint]{aastex}
\usepackage{emulateapj5}
\usepackage{epsf}
\usepackage{onecolfloat}
\newcommand{\Msun}{{\rm M}_{\sun}}

\begin{document}

\vspace{1mm}
\shortauthors{GNEDIN ET AL.}
\shorttitle{ADIABATIC CONTRACTION OF DARK MATTER HALOS}

\twocolumn[

\title{Response of dark matter halos to condensation of baryons:\\ 
cosmological simulations and improved adiabatic contraction model}

\author{Oleg Y. Gnedin\altaffilmark{1}, 
  Andrey V. Kravtsov\altaffilmark{2}, 
  Anatoly A. Klypin\altaffilmark{3},
  Daisuke Nagai\altaffilmark{2}
  }

\begin{abstract}
  The cooling of gas in the centers of dark matter halos is expected
  to lead to a more concentrated dark matter distribution.  The
  response of dark matter to the condensation of baryons is usually
  calculated using the model of adiabatic contraction, which assumes
  spherical symmetry and circular orbits.  In contrast, halos in the
  hierarchical structure formation scenarios grow via multiple violent
  mergers and accretion along filaments, and particle orbits in the
  halos are highly eccentric.  We study the effects of the cooling of
  gas in the inner regions of halos using high-resolution cosmological
  simulations which include gas dynamics, radiative cooling, and star
  formation.  We find that the dissipation of gas indeed increases the
  density of dark matter and steepens its radial profile in the inner
  regions of halos compared to the case without cooling.  For the
  first time, we test the adiabatic contraction model in cosmological
  simulations and find that the standard model systematically
  overpredicts the increase of dark matter density in the inner 5\% of
  the virial radius.  We show that the model can be improved by a
  simple modification of the assumed invariant from $M(r)r$ to
  $M(\bar{r})r$, where $r$ and $\bar{r}$ are the current and
  orbit-averaged particle positions.  This modification approximately
  accounts for orbital eccentricities of particles and reproduces simulation
  profiles to within $10-20\%$.  We present analytical fitting
  functions that accurately describe the transformation of the dark
  matter profile in the modified model and can be used for
  interpretation of observations.
\end{abstract}

 ]

\altaffiltext{1}{Space Telescope Science Institute,
       3700 San Martin Drive, Baltimore, MD 21218;
       {\tt ognedin@stsci.edu}}
\altaffiltext{2}{Dept. of Astronomy and Astrophysics,
       Kavli Institute for Cosmological Physics,
       The University of Chicago, Chicago, IL 60637;
       {\tt andrey,daisuke@oddjob.uchicago.edu}}
\altaffiltext{3}{Astronomy Department, New Mexico State University,
MSC 4500, P.O.Box 30001, Las Cruces, NM, 880003-8001;
       {\tt aklypin@nmsu.edu}}

\section{Introduction}
\label{sec:intro}
    
During the past decade dissipationless cosmological simulations have
shown that the density distribution within virialized halos of
different masses can be described by an approximately universal
profile \citep{dubinski_carlberg91,navarro_etal97,moore_etal98}.
Although non-baryonic dark matter exceeds baryonic matter by a factor
of $\Omega_{\rm dm}/\Omega_{\rm b} \approx 6$ on the average, the
gravitational field in the central regions of galaxies is dominated by
stars.  In the hierarchical galaxy formation model the stars are
formed in the condensations of cooling baryons in the halo center.  As
the baryons condense in the center, they pull the dark matter
particles inward thereby increasing their density in the central
region.

The response of dark matter to baryonic infall has traditionally been
calculated using the model of adiabatic contraction.
\citet*{eggen_etal62} were the first to use adiabatic invariants of
particle orbits to estimate the effect of a changing potential in a
contracting proto-galaxy.  \citet{zeldovich_etal80} used the adiabatic
invariant approach to calculate the contraction of lepton halos in
response to the cooling of baryons and set constraints on the
annihilation cross-section of leptons.  They also presented the first
analytical expressions for adiabatic contraction (AC) for purely
radial and circular orbits, as well as the numerical tests of such
model.  \citet{barnes_white84} showed that the reaction of the
particle orbits in spheroidal component to the slow growth of the disk
in their numerical experiments of disk and bulge evolution is indeed
adiabatic and can be described by a simple model that assumes circular
particle orbits and angular momentum conservation.

The present standard form of the AC model was introduced and tested
numerically by \citet[][see also \citeauthor{ryden_gunn87}
\citeyear{ryden_gunn87}]{blumenthal_etal86}.  This model assumes
spherical symmetry, homologous contraction\footnote{The halo can be
imagined consisting of spherical shells which contract in radius but
do not cross each other.}, circular particle orbits, and conservation
of the angular momentum: $M(r)r = {\rm const}$, where $M(r)$ is the
total mass enclosed within radius $r$.  With these assumptions, the
final dark matter distribution is calculated given the initial mass
profiles $M_{\rm dm}(r)$, $M_{\rm b}(r)$ and final baryon profile
$M_{\rm b}(r_f)$:
\begin{equation}
  \left[ M_{\rm dm}(r) + M_{\rm b}(r) \right] r =
  \left[ M_{\rm dm}(r) + M_{\rm b}(r_f) \right] r_f.
  \label{eq:standard}
\end{equation}
This model has been studied further by \citet{ryden88,ryden91} and
\citet{flores_etal93}. It is routinely used in mass modeling of
galaxies \citep[e.g.,][]{flores_etal93,dalcanton_etal97,mo_etal98,
courteau_rix99,vandenbosch01,vandenbosch_swaters01,klypin_etal02,seljak02} and
clusters of galaxies \citep[e.g.,][]{treu_koopmans02}.  The effect of
the contraction of the dark matter distribution is important for
studying star formation feedback on the centers of dark matter halos
\citep{gnedin_zhao02} and for comparing the abundance of dark matter
halos and galaxies as a function of circular velocity
\citep[e.g.,][]{gonzalez_etal00,kochanek_white01}. It is particularly
important in calculations of the dark matter annihilation signal from
the Galactic center \citep[e.g.,][]{gnedin_primack04,prada_etal04}.

Despite recent advances in numerical simulations, the model of
adiabatic contraction has never been tested in a cosmological context.
The tests performed to date \citep*{jesseit_etal02} assume spherical
symmetry and consider only the growth of a central concentration in an
isolated halo.  The hierarchical formation of halos is, in general,
considerably more complex than the simple picture of quiescent cooling
in a static spherical halo. Each halo is assembled via a series of
mergers of smaller halos, with the cooling of gas and contraction of
dark matter occurring separately in every progenitor.  The gas can be
re-heated by shocks during mergers and during accretion along the
surrounding filaments. Also, some objects may undergo dissipationless
mergers after the gas is exhausted or the cooling time becomes too
long. It was argued that dissipationless evolution erases the effect
of gas cooling on the DM distribution \citep{gao_etal04}.

In this paper we consider the effect of dissipation on the dark matter
distribution in high-resolution cosmological simulations. We also
present the first test of the AC model in the self-consistent
simulations of hierarchical structure formation and propose a simple
modification which describes numerical results more accurately.

\section{Cosmological simulations}
\label{sec:simulations}

We analyze high-resolution cosmological simulations of eight group and
cluster-sized and one galaxy-sized systems in a flat $\Lambda$CDM
model: $\Omega_{\rm m}=1-\Omega_{\Lambda}=0.3$, $\Omega_{\rm
b}=0.043$, $h=0.7$ and $\sigma_8=0.9$.  The simulations are performed
with the Adaptive Refinement Tree (ART) $N$-body$+$gasdynamics code
\citep*{kravtsov99, kravtsov_etal02}, an Eulerian code that uses
adaptive refinement in space and time and (non-adaptive) refinement in
mass to achieve the high dynamic range needed to resolve the halo
structure.

The cluster simulations have a peak resolution of $\approx 2.44h^{-1}$
kpc and DM particle mass of $2.7\times 10^{8}h^{-1}{\rm\ M_{\odot}}$
with only a region of $\sim 10h^{-1}\ \rm Mpc$ around each cluster
adaptively refined.  We analyze each cluster at a late epoch ($0 < z <
0.43$), when it appears most relaxed. This minimizes the noise
introduced by substructure on the azimuthally-averaged mass profiles.
The virial masses\footnote{We define the virial radius, $r_{\rm vir}$,
as the radius enclosing an average density of $180$ times the mean
density of the Universe at the analyzed epoch.}  of the clusters range
from $\approx 10^{13}h^{-1}{\ \rm M_{\odot}}$ to $3\times
10^{14}h^{-1}{\ \rm M_{\odot}}$.  

The galaxy formation simulation follows the early ($z \geq 4$)
evolution of a galaxy that becomes a Milky Way-sized object at $z=0$
in a periodic box of $6h^{-1}\ \rm Mpc$. The simulation is stopped at
$z \approx 3.3$ due to limited computational resources.  At $z=4$, the
galaxy already contains a large fraction of its final mass: $\approx 2
\times 10^{11}h^{-1}\ \rm M_{\odot}$ within $30 h^{-1}$ kpc.  The DM
particle mass is $9.18\times 10^5h^{-1}\ \rm M_{\odot}$ and the peak
resolution of the simulation is $183h^{-1}$~comoving pc.  This
simulation is presented in \citet{kravtsov03}, where more details can
be found.

For each halo, we analyze two sets of simulations which start from the
same initial conditions but include different physical processes.  The
first set of simulations follows the dynamics of gas
``adiabatically'', i.e. without radiative cooling.  The second set of
simulations (hereafter CSF) includes star formation, metal enrichment
and thermal supernovae feedback, metallicity- and density-dependent
cooling, and heating due to the extragalactic UV background.  Star
formation in the cluster simulations is implemented using the standard
Kennicutt's law and is allowed to proceed in regions with temperature
$T<10^4$K and gas density $n > 0.1\ \rm cm^{-3}$. In the galaxy
formation run, the star formation rate is proportional to the gas
density and stars are allowed to form at densities $n>50\ \rm cm^{-3}$. 
The difference in star formation prescriptions
in galaxy and cluster simulations, accounts for the difference
in spatial resolution. The prescription used in the galaxy
formation run is more appropriate when applied at the scale
of tens of parsecs \citep[see][for discussion]{kravtsov03}.

To identify dark matter halos we use a variant of the Bound Density
Maxima algorithm \citep{klypin_etal99}.  Dark matter particles in the
high-resolution region of the simulation are assigned a local density
calculated by using 24-particle SPH kernel.  We identify local density
peaks on a scale of $100h^{-1}$~kpc and analyze the density
distribution and velocities of the surrounding particles to test
whether a given peak corresponds to a gravitationally bound object. In
this study we only consider host halos: those that do not lie within a
larger virialized halo. We identify the center of each halo with the
position of the DM particle with the highest local density.  Based on
the convergence studies for the ART code
\citep{klypin_etal01,tasitsiomi_etal04}, we truncate the dark matter
profiles at the inner radius $4\Delta x_{\rm min}$, where $\Delta
x_{\rm min}$ is the smallest cell size: $2.44h^{-1}$ and $0.183h^{-1}$
comoving kpc in the cluster and galaxy formation runs, respectively.

\begin{figure}[t]
\centerline{\epsfysize3.5truein \epsffile{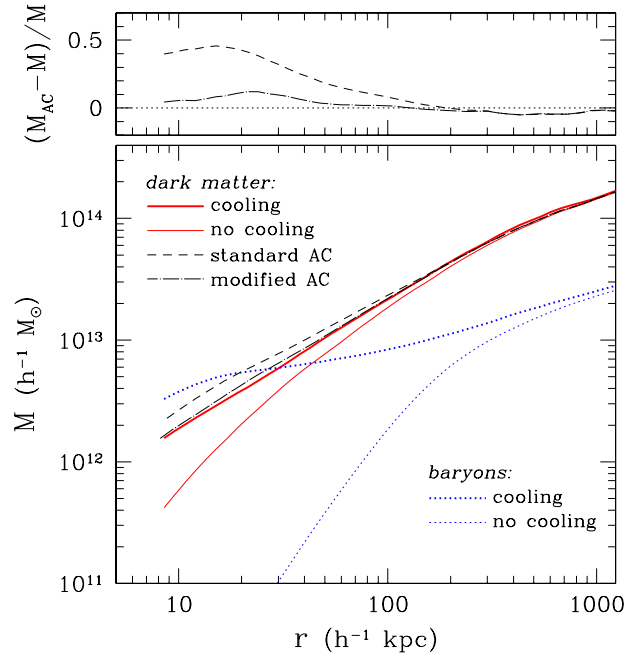}}
\caption{Mass profile of one of the clusters as a function of
physical radius.  The {\it solid} and {\it dotted} lines show the
profiles of dark matter and baryons (stars+gas) in the adiabatic ({\it
thin}) and cooling ({\it thick}) runs, respectively. The {\it dashed}
curve shows the prediction of the standard adiabatic contraction
model, while {\it dot-dashed} curve shows the improved model. The
profiles are truncated at four resolution elements of the simulation.
{\it Top panel:} relative mass difference between the adiabatic
contraction model and the DM profile in the CSF simulation. The {\it
dashed} line is prediction of the standard AC model, while {\it
dot-dashed} line shows our modified model.
  \label{fig:cl6_m}}
\end{figure}

\section{Effects of cooling on matter distribution}
\label{sec:cooling}

We start by comparing the spherically averaged distribution of baryons
and dark matter in the adiabatic and CSF simulations.  The comparison
reveals the effect of cooling and star formation because the two
simulation for each object have the same initial
conditions. Note that the cluster simulations likely suffer from the
``overcooling'' problem: the fraction of gas in the cold phase is
about a factor of $\sim 2-3$ higher than suggested by observations
\citep[e.g.,][]{balogh_etal01}.  The effect of cooling on mass
distribution is thus likely overestimated.  For the purposes of the
present study, however, this is acceptable.  In fact, the larger
effect of cooling allows us to emphasize the difference between the
simulations and the model.

\begin{figure}[t]
\centerline{\epsfysize3.5truein \epsffile{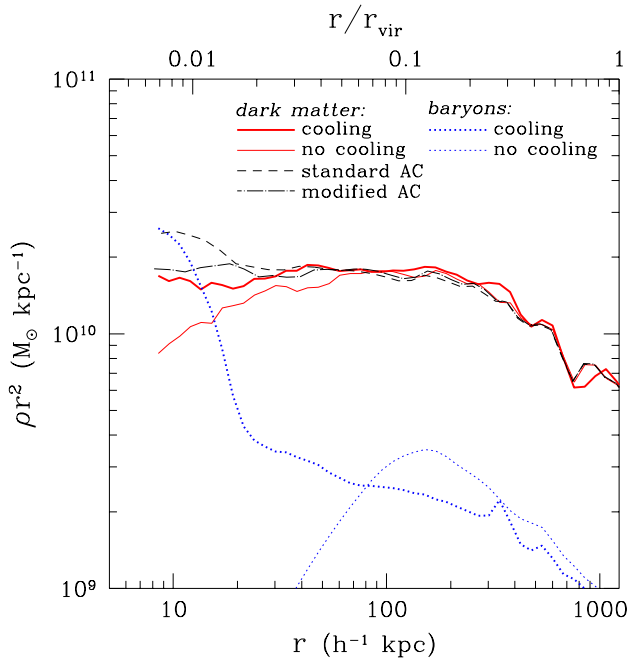}}
\caption{Density profile of the cluster shown in
  Figure~\ref{fig:cl6_m}, with the same line types.  In order to
  emphasize the differences at small radii, we plot the combination
  $\rho(r)r^2$ which is roughly constant for isothermal distributions.
  Physical radius is shown in $h^{-1}$ kpc ({\it bottom axis}) and as
  a fraction of the virial radius, $r_{\rm vir}$ ({\it top axis}).
  \label{fig:cl6_den}}
\end{figure}

The mass and density profiles of a representative cluster are shown in
Figures \ref{fig:cl6_m} and \ref{fig:cl6_den}.  These figures show
that cooling leads to an increase in the dark matter density within
$r\lesssim 50 h^{-1}$ kpc or $r/r_{\rm vir}\lesssim 0.04$ (see also
\citealt{tissera_domingueztenreiro98}).  The mass profile is affected
substantially at $r < 0.1 r_{\rm vir}$ and the change increases with
decreasing radius.  At larger radii the average mass profile is not
sensitive to baryon dissipation and the differences between the
adiabatic and CSF runs are the result of the slightly different
location of massive substructures.

\begin{figure}[t]
\centerline{\epsfysize3.5truein \epsffile{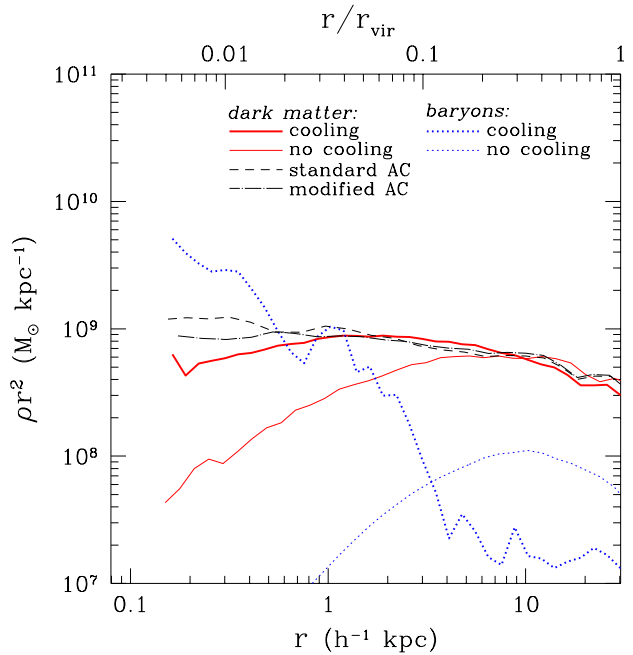}}
\caption{Density profile in the galaxy formation run at $z=4$ 
as a function of physical radius. Lines types are as in Figure~
\protect\ref{fig:cl6_m}.
  \label{fig:gal_den}}
\end{figure}

Figure~\ref{fig:gal_den} shows the density profiles in the galaxy
formation run at $z=4$. Qualitatively, the effect of cooling is
similar to that seen in the cluster simulations. In this case,
however, DM density in the CSF run is enhanced within a larger radius,
$r/r_{\rm vir} \lesssim 0.1$.  Also, baryons dominate the total
density at $r/r_{\rm vir} \lesssim 0.03$ in the galaxy simulation,
while in the cluster simulation they dominate only at $r/r_{\rm vir}
\lesssim 0.01$.  The difference is due to the considerably higher
fraction of cold ($T<10^4$~K) gas in the galaxy run: 80\% at $z=4$
versus $\sim 0.2-0.3$ in the cluster simulations.  These fractions
reflect a difference in densities, temperatures, and cooling times of
the gas in clusters and high-redshift galaxies.  Note that there are
differences in the mix of gas and stars.  At $z \lesssim 0.5$, most of
the baryon mass within the central regions of clusters is in the
stellar component of the cD galaxy.  In the galaxy run, on the other
hand, more than 80\% percent of baryons in the dense central disk are
still in the gaseous form.  The cluster and galaxy formation
simulations thus probe qualitatively different regimes of the
evolution of central baryon condensation.

\section{Testing the adiabatic contraction model}
\label{sec:acmodel}

\subsection{Standard Model}

In this section we test the standard prescription for adiabatic
contraction, given by equation~(\ref{eq:standard}).  In order to
calculate the model prediction for the final dark matter distribution
in the CSF run, we use the mass profile of baryons $M_{\rm b}(r_f)$
from this run and the mass profiles of dark matter and baryons from
the adiabatic simulation at the same epoch. We consider adiabatic
profiles as the initial profiles for the model, $M_{\rm dm}(r)$ and
$M_{\rm b}(r)$.  We then use equation~(\ref{eq:standard}) to predict
the DM distribution in the CSF run and compare the model prediction to
the actual dark matter profile in simulation.

The prediction of the standard AC model is shown in
Figures~\ref{fig:cl6_m}, \ref{fig:cl6_den}, and \ref{fig:gal_den} by
the dashed lines.  The fractional deviations of the model prediction
from the simulation for all analyzed systems are shown in
Figure~\ref{fig:cl_dm}. The standard model predicts the overall mass
enhancement but systematically overestimates its magnitude in the
inner regions, at $r/r_{\rm vir} \lesssim 0.1$.  This effect has
already been noticed in the original study by
\citet{blumenthal_etal86}.  Possible causes of the discrepancy could
be (1) non-spherical mass distribution and substructure; (2)
simultaneous evolution of the dark matter and baryonic components; and
(3) the assumption of circular orbits.  In the next section we
investigate whether the model can be improved by accounting for
orbital eccentricities.

\subsection{Modified model}
\label{sec:modified}

The orbits of particles in dark matter halos in simulations are highly
eccentric \citep[e.g.,][]{ghigna_etal98}.  The combination $M(r)r$ for
such orbits varies with the orbital phase and is not an adiabatic
invariant.  The conserved quantities for eccentric orbits are the
angular momentum, $J$, and the radial action,
\begin{equation}
  I_r \equiv {1\over\pi} \int_{r_p}^{r_a} v_r \, dr,
  \label{eq:ir}
\end{equation}
where $v_r$ is the radial velocity, and $r_p$ and $r_a$ are the peri-
and apo-center, respectively.  For non-crossing spherical shells, the
radial velocity can be expressed using the first integrals of motion
$C$ \citep[e.g.,][]{ryden_gunn87,gnedin_ostriker99} as
\begin{equation}
  v_r = \left( {2GM(r) \over r} - {J^2 \over r^2} + C \right)^{1/2}.
\end{equation}
The parameters $C$ change when shells cross, although the sum over all
shells is constant.  In the case of purely radial orbits in a
self-similar potential, we have $J=0$ and $r_p=0$, so that $I_r^2
\propto M(r_a) r_a$, and therefore the combination $M(r_a) r_a$ is
conserved during a slow change of the potential
\citep{blumenthal_etal86}.  The potentials of dark matter halos are,
in general, not self-similar while the angular momentum is not zero,
and the invariants cannot be reduced to a simple combination of $r$
and $M(r)$ that would be useful for predicting the final dark matter
profile.

It is interesting to check, though, if a combination that depends on
the average radius along the orbit, $\bar{r}$, rather than the
instantaneous value $r$ or the maximum $r_a$, is conserved better than
$M(r)r$ or $M(r_a) r_a$.  The orbit-averaged radius is
\begin{equation}
  \bar{r} = {2 \over T_r} \int_{r_p}^{r_a} r \, {dr \over v_r},
\end{equation}
where $T_r$ is the radial period.  

As a first check, we consider isochrone potential, $\Phi_{\rm iso}
\propto -[1+(1+(r/r_s)^2)^{1/2}]^{-1}$ \citep{binney_tremaine87}, for
which the radial action is known analytically as a function of $E$ and
$J$.  We choose the angular momentum that corresponds to realistic
eccentricities ($e \approx 0.7$): $J = J_c(E)/\sqrt{3}$, where
$J_c(E)$ is the angular momentum of a circular orbit of energy $E$.
We then check whether the combination $M(\bar{r}) \bar{r}$ is
proportional to $I_r^2$ for all orbits.  The ratio $M(\bar{r})
\bar{r}/I_r^2$ is constant at small radii ($r \ll r_s$) and large
radii ($r \gg r_s$) but varies in between by 50\%.  An alternative,
the apocenter ratio $M(r_a) r_a/I_r^2$ exhibits considerably larger
variation between the asymptotes, by about 300\%.  Thus the
combination $M(\bar{r}) \bar{r}$ is a better proxy to the invariant
than $M(r_a) r_a$.

In order to study more realistic orbits with the energies and angular
momenta relevant for cosmological simulations, we use a separate set
of high-resolution collisionless simulations of three Milky Way-sized
halos and one Virgo cluster-sized halo.  Galaxy-sized halos have about
one million DM particles within their virial radius \citep[see][for
details]{kravtsov_etal04}, while the cluster-sized halo has
approximately eight million particles \citep{tasitsiomi_etal04}.  All
runs have spatial resolution $\lesssim 10^{-3} \, r_{\rm vir}$.  We
approximate the potential of each halo with a spherically symmetric
NFW profile \citep{navarro_etal97} parameterized by the mass and
maximum circular velocity measured in the simulations.  We then
integrate the orbits in this potential starting with the particle
positions and velocities given by the simulations and calculate the
radial action (eq. \ref{eq:ir}) numerically.

\begin{figure}[t]
\centerline{\epsfysize4.5truein \epsffile{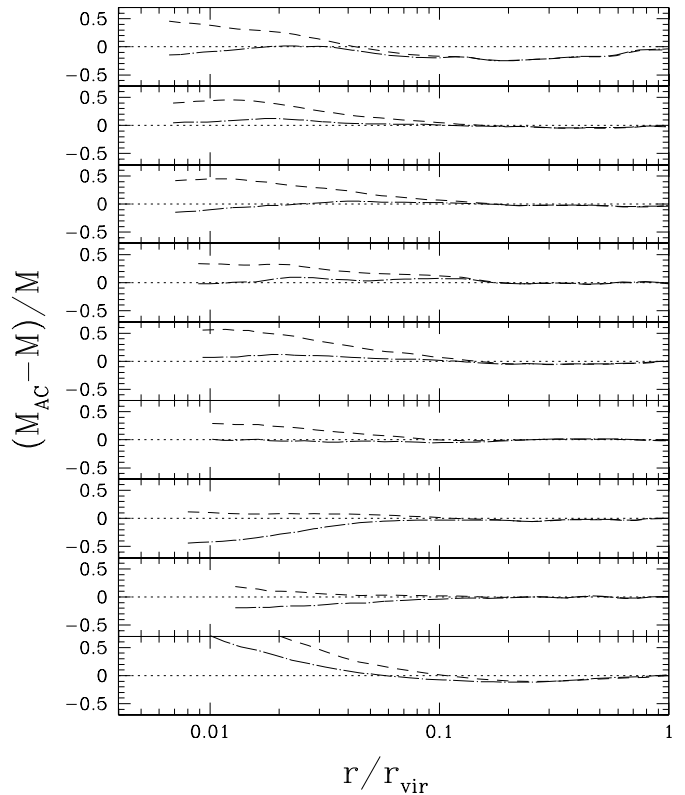}}
\caption{Fractional differences of the mass profiles predicted
  by the adiabatic
  contraction models and the simulation profiles for eight
  clusters ({\it top eight} panels) and one galaxy formation run (the
  {\it bottom} panel).  {\it Dashed} lines correspond to the standard
  model and {\it dot-dashed} lines show our modified model, eq.
  (\protect\ref{eq:modified}).  The cluster in the top panel is
  experiencing a merger event with a comparable mass cluster, which can
  be seen as an excursion of the profile at $r \sim 0.2 r_{\rm vir}$.
  \label{fig:cl_dm}}
\end{figure}

The orbits have a distribution of eccentricities that is very similar
to the isotropic distribution in analytical potentials studied by
\citet{vandenbosch_etal99}: the 20\%, 50\%, and 80\% quartiles are $e
= 0.39, 0.61, 0.79$, respectively.  Given such a wide eccentricity
distribution, the orbit-averaged radius $\bar{r}$ varies for particles
at a given current radius $r$ depending on the orbital phase.
Nevertheless, the mean relation at $10^{-3}\lesssim r/r_{\rm
vir}\lesssim 1$ can be described by a power law function
\begin{equation}
  {\bar{x}}=A x^{w}, \quad x\equiv r/r_{\rm vir},
  \label{eq:rrave}
\end{equation}
with small variations in the parameters $A$ and $w$ from halo to halo
and from epoch to epoch.  The mean values are $A\approx 0.85 \pm 0.05$
and $w \approx 0.8 \pm 0.02$, which we use as our fiducial parameters.
This power-law dependence reflects typical energy and eccentricity
distributions of particles in cold dark matter halos.  At $x < 0.44$
the average radius is larger than the current radius, while at $x >
0.44$ it is smaller.

The distribution of eccentricities at a current radius $r$ leads also
to a distribution of each combination $M(r)r$, $M(r_a)r_a$, and
$M(\bar{r})\bar{r}$.  None of these combinations is proportional to
the invariant $I_r^2$ for individual orbits.  What we are looking for
is a proxy to the invariant for an ensemble of orbits representing a
spherical shell.  Such a proxy would allow us to calculate the
transformation of the dark matter profile imagined of consisting of
spherical shells.  Therefore, instead of individual particle orbits we
look at the average ratios $M(r_a)r_a/I_r^2$ and
$M(\bar{r})\bar{r}/I_r^2$ in radial bins.  Across all bins the
average ratio $M(r_a)r_a/I_r^2$ varies by a factor 3.4, while the
ratio $M(\bar{r})\bar{r}/I_r^2$ varies only by a factor 2.2.
Therefore, our insight gained from analytic orbits in isochrone
potential is still valid for isotropic orbits in NFW potential: the
combination $M(\bar{r})\bar{r}$ is a better proxy to the invariant.
Moreover, we have found that the mixed combination $M(\bar{r})r/I_r^2$
varies even somewhat less, by a factor of 2 across all radii.  This
latter combination thus is our best proxy for the radial action.

Motivated by these considerations, we propose a modified adiabatic
contraction model based on conservation of the product of the current
radius and the mass enclosed within the orbit-averaged radius:
\begin{equation}
  M(\bar{r})r= {\rm const}.
  \label{eq:modified}
\end{equation}
Using equation (\ref{eq:rrave}) we compare
our gasdynamic simulations with this modified model.

Figures \ref{fig:cl6_m}--\ref{fig:cl_dm} show that the modified model
provides a more accurate description
  of the simulation
results than the standard model for most of the objects. Although
there are still deviations of the model and simulation profiles, there
is no systematic over- or under-prediction of the mass for all
objects.  Typical deviations are $\lesssim 10\%$. Note that for the
high-redshift galaxy disk both the standard and modified AC models
overpredict the mass profile significantly, although the modified
model is still closer to the simulated profile.  The discrepancy is
large in the inner kiloparsec where the most of the mass is in the
thin gaseous disk.  The adiabatic contraction model cannot be applied
in such regime and further numerical simulations are needed to
investigate the dark matter distribution in the galactic center.

An interesting question is why the adiabatic contraction approximation
works as well as it does?  Although the model assumes that gas cooling
affects DM distribution in the final object adiabatically,
the particles experience contraction in separate unconnected halos and
undergo some degree of violent relaxation in subsequent mergers.  To
check the evolution of individual particle orbits we have compared
their physical distance $r$ to the center of the main progenitor and
quantities $M(r)r$, $M(\bar{r})r$ at a number of epochs from $z=0$ to
$z=4$ for one of the clusters. There is a substantial scatter in the
relation of these quantities between the present and higher redshifts.
However, some interesting trends for the {\it averages} of $M(r)r$ and
$M(\bar{r})r$ can be observed. First of all, we find that both radii
and products of radii and mass evolve as the object grows
hierarchically, especially during major mergers. However, between
mergers, during the periods when a substantial amount of gas cools,
$M(r)r$ and $M(\bar{r})r$ seem to be conserved. The latter combination
is conserved better than the former, which may explain why our
modification to the AC model is successful.

We find also that most ($\approx 70\%$) particles within the central
$25h^{-1}$~kpc of the cluster center at $z=0$ come from a single
progenitor at $z=1$.  This progenitor is not the most massive but has
the highest central density.  When it merges with a more massive
progenitor at $z\approx 0.5$ to form the final system, its particles
are most tightly bound and end up dominating the mass in the central
region of the merger remnant.  This is consistent with other merger
simulations which show that particles from the progenitor with the
densest central region dominate the inner regions of the remnant
\citep{boylankolchin_ma04}.

\begin{figure}[t]
\centerline{\epsfysize3.5truein \epsffile{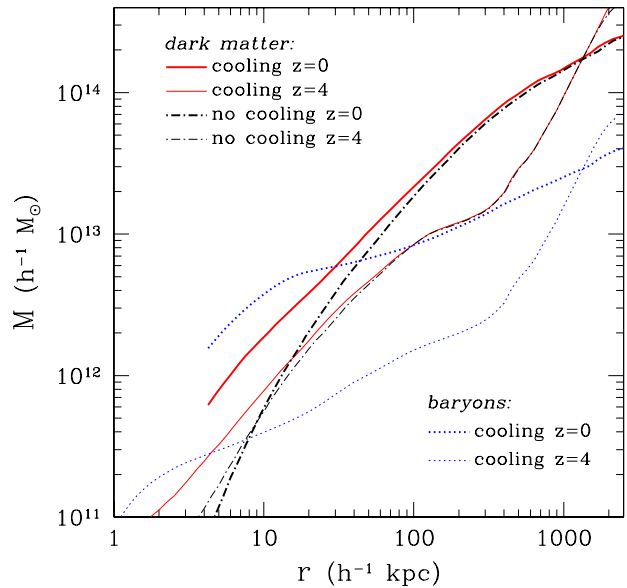}}
\caption{The effect of gas cooling on the mass profile at different
  epochs for the cluster shown in Figures~\ref{fig:cl6_m} and
  \ref{fig:cl6_den}. {\it Thin} and {\it thick} lines show the
  profiles at $z=4$ and $z=0.18$, respectively: {\it dotted} lines ---
  profiles for baryons (gas$+$stars) in the CSF run; {\it dot-dashed}
  lines --- profiles of DM in the adiabatic run; {\it solid} lines ---
  DM profiles in the CSF run.
  \label{fig:cl6_mz}}
\end{figure}

\section{How steep is the central dark matter profile?}
\label{sec:centralpro}

Dark matter distribution in the central regions of galaxies and
clusters is important for testing the CDM paradigm and interpreting
the observations. It is generally believed that dissipation by baryons
would steepen the dark matter profile
\citep{barnes_white84,blumenthal_etal86,ryden_gunn87}. However,
\cite{loeb_peebles03} and \citet[][hereafter G04]{gao_etal04} recently
suggested that after an early epoch of cooling and rapid star
formation, subsequent dissipationless mergers would erase the
cooling-induced central concentration of dark matter.  They put forth
a hypothesis that the NFW profile is a dynamical attractor, in the
sense that remnants of dissipationless mergers are driven to the
central profile with the cusp $\rho(r)\propto r^{-1}$, even if their
progenitors have steeper DM profiles. As a supporting argument G04 use
observation that the density in the inner regions of halos in {\it
dissipationless} simulations is approximately constant after the
initial period of rapid mass growth.  The attractor hypothesis
together with observation that stars dominate gravitationally in the
centers of galaxies leads to the conclusion that the dark matter
profile in galaxies which experienced dissipationless mergers should
be shallower than $r^{-1}$.

The results of our simulations do not support this conclusion.  Given
the importance of the problem, it is worth discussing the differences
between the simulations and analysis of G04.  Figure \ref{fig:cl6_mz}
compares the mass profile of the cluster shown in
Fig.~\ref{fig:cl6_m} at $z=0$ with its most massive progenitor at
$z=4$.  We first verify that in adiabatic simulations the DM mass
profile in the inner $\approx 10h^{-1}$~kpc is approximately constant
from $z=4$ to the present epoch, in agreement with analyses of
\cite{fukushige_makino01} and G04.  Within the physical radius $r =
10h^{-1}$~kpc the enclosed mass is $M(r)\approx 5\times 10^{11}\,
h^{-1}\, \Msun$ at both epochs.

The evolution is very different in the run with cooling and star
formation.  By $z=4$, a considerable stellar mass ($\sim 3\times
10^{11}\, h^{-1}\, \Msun$) has formed within $10\, h^{-1}$ kpc of the
center of the main cluster progenitor.  However, the baryon mass
within central $10\, h^{-1}$ kpc continues to grow and increases by a
factor of ten between $z=4$ and $z=0.2$. Approximately 50\% and 70\%
of those stars form at $z<1$ and $z<2$,
respectively. Note that the stars and cold gas in cluster cores are
accumulated both due to direct cooling in the core and via accretion
during mergers \citep{motl_etal04}.

As can be seen in Figure~\ref{fig:cl6_mz}, such substantial increase
in the baryon mass leads to the increase of the DM density in the
inner regions.  The final dark matter mass within $10\, h^{-1}$ kpc is
$\approx 2\times 10^{12}\, h^{-1}\, \Msun$, or a factor of four larger
than the mass in the adiabatic simulation.  Therefore, one of the
major differences between our simulations and analysis of G04 is that
in the simulation the density of both baryons and dark matter
increases at lower redshifts, while G04 assume that the density of DM
decreases as the total mass distribution is driven to the NFW profile.
Note also that G04 assume that star formation and cooling in the
centers of massive ellipticals effectively stops at $z\sim 2-3$, while
in our simulations cooling and star formation continue at lower
redshifts with more than half of the stars formed at $z<2$. This can
be a deficiency of simulations which, as we mentioned above, suffer
from the overcooling problem. The fraction of baryons in the cold gas
and stars within the virial radius of clusters at $z=0$ in our
simulations is in the range $\sim 0.3-0.4$, at least a factor of two
higher than observed for the systems of the mass range we consider
\citep*{lin_etal03}. The overcooling and relatively late star formation
is a generic problem of cosmological simulations and is hardly
realistic. It likely indicates that some mechanism suppressing
cooling is needed.  

If we follow G04 and assume that cooling ceases at late epochs, the
question is then whether subsequent dissipationless evolution erases
the prior effect of cooling on the concentration of DM and drives the
overall stellar$+$DM profile to the NFW form, as required by the
attractor hypothesis.  Several recent studies have considered the
effect of major mergers and dynamical effects of substructure on the
dark matter profiles in the inner regions of halos
\citep{dekel_etal03,dekel_etal03b,elzant_etal04,boylankolchin_ma04,ma04,
nipoti_etal04}.  Dissipation of the orbital energy of massive subhalos
by dynamical friction can heat the dark matter particles of the
host and reduce the DM density in the inner regions.  On the other
hand, DM particles of the sinking subhalos replace the host
particles and the overall central density profile stays approximately
constant or even becomes steeper, depending on internal structure,
spatial distribution, and orbital parameters of subhalos \citep{ma04}.

Major mergers can lead to a more drastic and violent rearrangement of
matter in halos compared to the effects of substructure.
\citet{boylankolchin_ma04} present a set of merger simulations, which
are most relevant for our discussion. They show that mergers of halos
with constant density cores produce a remnant with a constant density
core, but mergers of the cored and cuspy halos produce a cuspy remnant.
Their analysis shows that the initial cusp is remarkably stable and
the density distribution of the merger remnants retains memory of the
density profiles of their progenitors. This is in good
agreement with the analysis of merger experiments presented by
\citet{kazantzidis_etal04}. The merger remnant of the DM halos with a
steep inner density profile ($\rho(r)\propto r^{-1.8}$) retains the
initial slope of the inner cusp. The mergers of halos with embedded
stellar disks also produce DM halos with the inner cusp not shallower
than the initial. 

Finally, to test the effect of dissipationless mergers on the DM
density profile and the validity of the AC model, we repeated the
cluster simulation shown in Figure~\ref{fig:cl6_mz} with the cooling
turned off at $z<2$.  The evolution is thus dissipationless at low
redshift.  The cluster undergoes a number of minor mergers, as well as
one approximately equal-mass merger at $z\sim 0.5$. The amount of cold
gas and stars within the virial radius of this cluster at $z=0$ is
$\approx 14\%$, a factor of two smaller than in the original
simulation and similar to the stellar fractions observed in galaxy
clusters \citep[e.g.,][]{lin_etal03}.  We find that although the
effect of contraction in this case is smaller, as less baryons
condense in the center, the dark matter profile is still steeper than
in the adiabatic run and is well described by the adiabatic
contraction model.  Dissipationless mergers that the main cluster
progenitor has undergone at $z<2$ apparently have not erased the
steepening of the profile due to the cooling.

The attractor hypothesis is thus not supported both by dissipationless
merger simulations and our simulation in which cooling is stopped at high
redshift.  The effects of dissipation on the DM distribution
in the progenitors is retained, at least to a certain extent, in the
density distribution of their descendant.  The overall effect of
merging is just to mix dark matter particles within the same
distribution, while baryon dissipation leads to a significant increase
of the dark matter density.

Finally we note that for the halos that harbor a supermassive black
hole the density distribution within its sphere of influence
(typically $10-100$ pc) is determined by the interaction between the
black hole, stars, and DM \citep[e.g.,][]{gnedin_primack04}.

\section{Discussion and Conclusions}
\label{sec:discussion}

We have analyzed results of self-consistent cosmological simulations
of eight galaxy clusters and one galactic halo with and without
cooling and star formation. The comparison of adiabatic and CSF
simulations shows that cooling increases the total density and the
density of dark matter at $r\lesssim 0.1r_{\rm vir}$. This agrees
qualitatively with results of other recent simulations
\citep{tissera_domingueztenreiro98,lewis_etal00,pearce_etal00,valdarnini02}.
Note that at $r \gtrsim 0.01 r_{\rm vir}$ modern dissipationless
simulations have reached a robust convergence
\citep[e.g.,][]{diemand_etal04}.  We conclude therefore that further
progress in making predictions for the DM distribution on small scales
requires studying gas dissipation.  The effect of cooling also needs
to be taken into account when comparing the simulations with
observations.

We have presented the first tests of the adiabatic contraction model
in self-consistent high-resolution cosmological
simulations.\footnote{After this paper was completed we have learned
  of a different study of the AC model in cosmological simulations
  (Gottbrath \& Steinmetz, 2000 unpublished). These authors found that
  the AC model works adequately at the resolved scales. However, the
  resolution of the simulations used in their study is considerably
  lower than resolution of our simulations, and the scales where we
  find significant discrepancy between AC model and simulations were
  not resolved.} We find that the standard AC model systematically
overpredicts the increase of the dark matter density in the inner
$\lesssim 0.05r_{\rm vir}$.  We have shown that the model can be
improved by a simple modification of the assumed conserved invariant
from $M(r)r$ to $M(\bar{r})r$, where $r$ and $\bar{r}$ are the current
and orbit-averaged particle positions. This modification approximately
accounts for the eccentricity of particle orbits.  Our improved model
describes profiles in simulations considerably better than the
standard model, with the average accuracy of $10-20\%$.

\citet{jesseit_etal02} have used controlled simulations of isolated
spherical halos with a growing central concentration of baryons to
test the validity of the standard AC model. They find that generally
the standard model describes their simulation more accurately than we
find in our tests against cosmological simulations.  Possible causes
of the discrepancy with our results are the differences of (i) orbital
distributions for the isolated halos versus cosmological halos formed
by mergers; and (ii) formation histories and degree of violent
relaxation.  In particular, \citet{jesseit_etal02} find that the
standard model significantly overpredicts the contraction effect when
the growth of the central baryon concentration is rapid compared to
the dynamical time of the halo.

In order to understand this discrepancy, we have run isolated
simulations of a live NFW halo with an analytical contracting disk
using the publicly available tree code GADGET \citep{gadget01} with
$10^6$ particles.  In agreement with \citet{jesseit_etal02}, we find
that both the standard and the modified models work well (within 20\%)
in the regions where the baryon density is comparable to the final
dark matter density.  However, the modified model becomes
progressively more accurate as the baryons dominate over the dark
matter by a factor of 2-3.  In this case the ratio of the baryon
density to the initial DM density is a factor of $\sim 20-50$, while
the ratio of the final to initial DM density (compression ratio) is
$\sim 10$.  For example, the modified model should apply in the
central parts of the Galaxy.  From these experiments we conclude that
the accuracy of our model is determined both by the orbital
distributions and by the amount of compression.

Given that dark matter halos assemble via mergers and violent
relaxation, it is somewhat of a puzzle that the adiabatic contraction
model reproduces the results of simulations so well.  The success of
the model seems to imply that the effect of the central baryon
condensation on the dark matter distribution is independent of the way
in which this condensation is assembled.  At the same time, it may
simply be due to the fact that the central region is dominated by
particles from a single densest progenitor.  If the progenitor halo
contracts in response to the cooling of baryons early on and then
approximately preserves the shape of its inner density profile during
subsequent mergers, as suggested by merger simulations (see
\S~\ref{sec:centralpro}), the AC model applied to the final mass
distribution is expected to work.

In Appendix we provide analytical fitting functions that describe the
contraction of an initial NFW profile in our modified model.  These
functions can aid in interpretation of observations of galaxy halos
and clusters of galaxies.  We show that the inner slope of the dark
matter density profile $\gamma$ is determined by the shape of the
baryon profile (see eq.~[\ref{eq:gamma}]).  For the specific cases of
the exponential disk and Hernquist model, both the baryon and the
contracted DM profiles have the asymptotic slope $\gamma=1$.

Our results have several implications for the efforts to test
predictions of the CDM model observationally. The test that received
much attention in the last decade is the density distribution in the
inner regions of galaxies and clusters. Extensive convergence studies
have shown that modern highest-resolution dissipationless simulations
agree in their predictions: the average logarithmic slope of the
density profile at $r=0.01r_{\rm vir}$ is $\gamma \approx 1.3$ with a
substantial scatter of $\pm 0.3$ from object to object
\citep{fukushige_etal04,tasitsiomi_etal04,navarro_etal04,reed_etal04,
diemand_etal04}.  At the same time, despite a significant decrease in
the smallest reliably resolved scale, the logarithmic slope continues
to get shallower with decreasing radius without reaching an asymptotic
value.  For the purposes of the present discussion, it suffices that
dissipationless simulations have converged at the scales where the
effects of dissipation become important.

Observational measurements of the dark matter density distribution are
notoriously difficult. Several approaches have been used but in each
case opposite conclusions are reached by different researchers, often
after analyzing the same data. The rotation curves of dark matter
dominated dwarf and low-surface brightness galaxies tend to favor
density profiles shallower than predicted by CDM
\citep[e.g.,][]{simon_etal03,deblok_etal03}. The results, however, are
sensitive to the resolution of rotation curves, presence of bulges and
non-circular motions, and for many galaxies the profiles can be
reconciled with theoretical expectations
\cite[e.g.,][]{swaters_etal03,rhee_etal04}
\cite[but see][]{deblok03}.  The analysis of density
distribution for bright galaxies is complicated by the uncertain
contribution of stars to the total mass profile
\citep[e.g.,][]{treu_koopmans02,mamon_lokas04}.  Some analyses tend to
favor inner slopes shallower than predicted by CDM
\citep[e.g.,][]{gentile_etal04}, but others deduce slopes of the inner
profiles (at least marginally) consistent with predictions
\citep{treu_koopmans02,treu_koopmans04,koopmans_treu03,jimenez_etal03}.

The density distribution in galaxy clusters can, in principle, provide
a cleaner test of the models because effects of ``gastrophysics'' on
the DM distribution are expected to be smaller and simpler.  The
advances in lensing analyses, X-ray observations, and high-resolution
spectroscopy allow observers to obtain constraints on the DM
distribution using a variety of techniques.  Recent observational
analyses using strong lensing \citep[e.g.,][]{tyson_etal98},
high-resolution X-ray imaging \citep[][but see
\citeauthor{david_etal01} \citeyear{david_etal01} and
\citeauthor{arabadjis_etal02}
\citeyear{arabadjis_etal02}]{ettori_etal02,katayama_hayashida04}, and
lensing$+$velocity dispersion measurements for the central galaxy
\citep{sand_etal02,sand_etal04} seem to favor shallow ($\gamma
\lesssim 0.5-1$) inner density distributions.  If these
 observational results stand, it would be a major
problem for the CDM because dissipation generally makes the
discrepancy worse.

Many of the systematic effects and validity of the key assumptions in
such measurements are yet to be explored. For example, deviations from
spherical symmetry in the mass distribution allows for the inner
slopes of $\gamma \gtrsim 1$
\citep{dalal_keeton04,bartelmann_meneghetti04}.  Some of our results
also call into question the assumptions used to derive observational
constraints. For example, it is often assumed that at small scales the
DM profile can be well approximated by a power law, $\rho(r)\propto
r^{-\gamma}$.  It is also assumed that dissipation would steepen the
profile predicted from dissipationless simulations while retaining its
power law form.  As we show in Appendix, however, the steepening of
the profile due to cooling is in general scale-dependent.  For
realistic cases, the profiles of dark matter and baryons at $r
\lesssim 0.01 r_{\rm vir}$ should be quite similar. These scales are
exactly where the profiles of massive ellipticals and central cluster
galaxies are probed by spectroscopic measurements in observations. If
incorrect assumptions are made about the DM distribution, there is a
danger of oversubtracting the contribution of baryons to the total
profile.  Stars dominate the inner region and only a small
overestimate of the stellar mass-to-light ratio could lead to a much
lower residual density of dark matter.  

The fact that the standard AC model overpredicts the effect of cooling
on mass distribution means that observational analyses that use the
standard model \citep[e.g.,][]{treu_koopmans02,treu_koopmans04}
provide somewhat less stringent limits on $\gamma$ than claimed.  Our
results for the disk galaxy at high redshift show that even the
improved model can still overestimate the effect in the inner region
of the gaseous disk.  Extrapolation of the model to very small radii,
as is often done in predictions of the dark matter annihilation
signal, can therefore be dangerous.  Further high-resolution
gasdynamics simulations are needed to probe the effect of cooling in
the centers of dark matter halos.

\acknowledgements 

We would like to thank Frank van den Bosch, Leon Koopmans, Joel
Primack, David Weinberg, and Simon White for discussions and useful
comments on the manuscript, and Andrew Zentner and Stelios Kazantzidis
for discussions and analysis of the density profiles in the controlled
merger experiments.  We are grateful to Nicole Papa for careful
reading of the manuscript.  We would also like to acknowledge Bocage
of San Saba Vineyards.  This work was supported by the National
Science Foundation under grants AST-0206216 and AST-0239759 to the
University of Chicago, by NASA through grant NAG5-13274, and in part
by grant by the Kavli Institute for Cosmological Physics (KICP), an
NSF Physics Frontier Center, through grant NSF PHY-0114422.  O.Y.G. is
supported by the STScI Fellowship.  D.N. is supported by the NASA
Graduate Student Researchers Program and by NASA LTSA grant NAG5-7986.
The simulations presented here were performed on the IBM SP4 system
({\tt copper}) of the National Computational Science Alliance.

\onecolumn
\appendix
\section{Analytical fits for the contracted dark matter mass profile}
\label{sec:fits}

In many applications it can be useful to have an analytical formulae to
estimate the compression of dark matter due to baryonic condensation.  We
find that a simple procedure described below provides an accurate fit
for the compression of an initial NFW profile.

We assume that the initial distributions of dark matter and baryons
are both given by the NFW profile, $M_i(r)$, with a concentration
parameter $c$. Subsequently the baryons cool and form stars and their
final profile is given by $M_{\rm b}(r_f)$, with the ratio of the
baryon-to-total mass at the virial radius $M_{\rm b}(r_{\rm
  vir})/M_{\rm vir} = f_{\rm b}$.  This ratio does not need to equal
the universal baryon fraction and may deviate from it depending on the
details of hierarchical formation and heating by the extragalactic UV
flux.  We consider two representative examples of the final baryon
distribution: an exponential disk, which is appropriate for spiral
galaxies, and a Hernquist model, which adequately describes the
stellar profiles in elliptical galaxies.  Both of these profiles are
characterized by a scalelength $r_{\rm b}$.

With the usual assumption of spherical homologous contraction, our
goal is to calculate the final radius $r_f$ of the dark matter
particles initially located at $r > r_f$.  It is useful to define the
contraction factor $y \equiv r_f/r$.  The equation we need to solve
for $y$ is
\begin{equation}
  r M_i(\bar{r}) = 
      \left[ (1-f_{\rm b}) M_i(\bar{r}) + M_{\rm b}(\bar{r}_f)
      \right] r_f.
\end{equation}
Let us divide both sides of the equation by
$M_i(\bar{r}) r_f$ and use dimensionless radius $x \equiv
r/r_{\rm vir}$ and mass $m \equiv M/M_{\rm vir}$:
\begin{equation}
  {1 \over y} = 1 - f_{\rm b} + 
                {m_{\rm b}(\overline{xy}) \over m_i(\bar{x})}.
  \label{eq:y}
\end{equation}
By definition, there is no contraction at the virial radius: $y(1)=1$.

An important simplifying feature of the profiles we consider
here (NFW, exponential disk, Hernquist model) is that at $x
\ll 1$ the enclosed mass of all three profiles grows with radius as
$m(x) \propto x^2$.  The first two terms of the Taylor expansion at $x
\ll 1$ are:
\begin{eqnarray}
  m_i(x) & = & g_c \left[ \log{(1+cx)} - {cx \over 1+cx} \right]
         \approx x^2 {g_c c^2 \over 2} \left( 1 - {4c\over 3}x +... \right)
         \nonumber\\
  m_{\rm b}^{H}(x) & = & f_{\rm b} x^2 \left( 1+r_{\rm b} \over x+r_{\rm b} 
         \right)^2 
         \approx x^2 {f_{\rm b} (1+r_{\rm b})^2 \over r_{\rm b}^2}
         \left( 1 - {2 \over r_{\rm b}}x +... \right) \\
  m_{\rm b}^{E}(x) & = & f_{\rm b} \left[ 1 - \left(1+{x \over r_{\rm b}}\right)
         e^{-x/r_{\rm b}} \right]
         \approx x^2 {f_{\rm b} \over 2 r_{\rm b}^2}
         \left( 1 - {2 \over 3 r_{\rm b}}x +... \right),
         \nonumber
\end{eqnarray}
where $g_c \equiv [\log{(1+c)} - c/(1+c)]^{-1}$.  Since our fit for
the orbit-averaged radius $\bar{x}$ is also a power-law ($\bar{x} = A
\, x^{w}$, $w = 0.8$) the mass ratio in eq. (\ref{eq:y}) is constant
at $x \ll r_{\rm b},c^{-1}$, and therefore $y(x)$ is approximately
constant.

The finite value $y_0 \equiv y(0)$ is pivotal to constructing the
fitting function for $y(x)$.  We obtain the following equation for
$y_0$ by retaining only the first terms of the Taylor expansion:
\begin{equation}
  {1 \over y_0} = 1 - f_{\rm b} + a \, y_0^{2w},
  \label{eq:y0}
\end{equation}
where $a = 2 f_{\rm b} (1+r_{\rm b})^2 / (r_{\rm b}^2 g_c c^2)$ for
the Hernquist model and $a = f_{\rm b} / (r_{\rm b}^2 g_c c^2)$ for
the exponential disk.  There is no analytical solution for this
algebraic equation, but we can obtain a very accurate approximate
solution as follows.  We first consider the standard model $w=1$,
for which eq. (\ref{eq:y0}) becomes a cubic equation.  The solution is
\begin{equation}
  y_{(1)} = \left( Q^{1/2} + {1 \over 2a} \right)^{1/3} -
            \left( Q^{1/2} - {1 \over 2a} \right)^{1/3}, \quad
  Q \equiv \left( {1-f_{\rm b} \over 3a} \right)^3 + {1 \over (2a)^2}.
  \label{eq:y3sol}
\end{equation}
There is only one real solution because $Q$ is always positive.  We
then note that eq. (\ref{eq:y0}) with $w = 0.8$ differs from the
cubic equation only at $a \gtrsim 1$.  We can therefore obtain a
solution for $y_0$ by matching the solution of the cubic equation for
small $a$ with the correct asymptote for large $a$.  We find that the
following modification produces the right asymptotic solution for $a >
1$:
\begin{equation}
  y_{(w)} = \left( \hat{Q}^{1/2} + {1 \over 2a} \right)^{1/(1+2w)} -
            \left( \hat{Q}^{1/2} - {1 \over 2a} \right)^{1/(1+2w)}, \quad
  \hat{Q} \equiv {1\over a^3} \left( {1-f_{\rm b} \over 1+2w} 
            \right)^{1+2w}
           + {1 \over (2a)^2}.
  \label{eq:y26sol}
\end{equation}
Finally, we link the two asymptotes with a weight function that
we empirically found to minimize the error of the fit:
\begin{equation}
  y_0 = y_{(1)} \, e^{-2a} + y_{(w)} \, \left( 1 - e^{-2a} \right).
  \label{eq:y0sol}
\end{equation}
The relative error of this fit, compared with the direct numerical
solution, is only 0.2\% on the average for all $a$ and $f_{\rm b}$.
The asymptotes $a \ll 1$ and $a \gg 1$ are reproduced almost exactly,
while the maximum error of 0.6\% occurs at $a \sim 1$.

Now we need to connect the boundary values $y(0) = y_0$ and $y(1) = 1$
by an interpolating function.  This can be done very accurately with
the following trick.  Let us define an auxiliary function
\begin{equation}
  t(x,y) \equiv \left[ 1 - f_{\rm b} +
    {m_{\rm b}(\overline{xy}) \over m_i(\bar{x})} \right]^{-1}.
  \label{eq:t}
\end{equation}
It satisfies the boundary conditions $t(0,y_0) = y_0$ and $t(1,1) =
1$.  Since $y(x)$ varies slowly near the two bounds, we can use the
function $t(x,y)$ to calculate the asymptotic solutions: $y(x \ll 1)
\approx t(x,y_0)$ and $y(x \approx 1) \approx t(x,1)$.  The two
asymptotes can be linked by a smooth weight function:
\begin{equation}
  y(x) = t(x,y_0) \, e^{-bx} +
         t(x,1) \, \left( 1 - e^{-bx} \right).
  \label{eq:ysol}
\end{equation}
The exponent $b$ that minimizes the error of the fit can be found by
an approximate Taylor expansion of eq. (\ref{eq:y}) near $x=0$.
We find
\begin{equation}
  b = {2 y_0 \over 1-y_0} \left({2 \over n r_{\rm b}} - {4c \over 3}\right)
      \left( 2.6 + {1-f_{\rm b} \over a y_0^{2w}} \right)^{-1},
  \label{eq:b}
\end{equation}
where $n=1$ for the Hernquist model and $n=3$ for the exponential
disk.  We have tested the accuracy of eq. (\ref{eq:ysol}) against
direct numerical solution for $y(x)$ varying the parameters in the
range $4 < c < 20$, $0.02 < f_{\rm b} < 0.24$, $0.01 < r_{\rm b} <
0.07$, which should cover most of the cases of interest.  We find an
average relative error of 2\% for the Hernquist model and 1\% for the
exponential disk.  The maximum error is 7\% and 4\%, respectively, and
occurs at $x \sim r_{\rm b}$.

\begin{figure}[t]
\centerline{\epsfysize3.5truein \epsffile{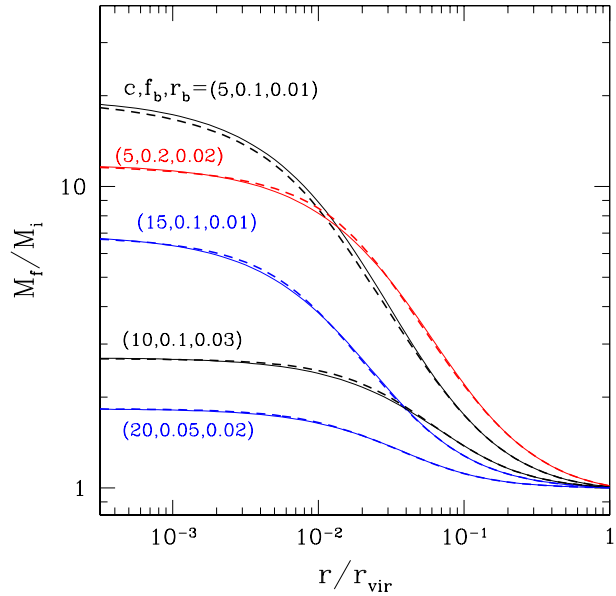}}
\caption{Increase of the dark matter mass
  for the initial NFW models assuming the exponential disk for the
  final baryon profile.  {\it Solid} lines are the numerical solution,
  while {\it dashed} lines are our fitting formulae. Five sets of
  lines show that our analytical fit (eq. [\protect\ref{eq:fm}])
  reproduces the direct numerical solution for various combinations of
  the parameters $c$, $f_{\rm b}$, and $r_{\rm b}$, with a typical 
  accuracy of better than a few percent.
  \label{fig:fit_nfw}}
\end{figure}

We can now easily find the final dark matter profile at the contracted
radius $x_f = x y(x)$: $M_{\rm dm,f}(xy) = M_{\rm dm,i}(x)$.  If
desired, the profile can be re-mapped to the grid of initial radii $x$
by interpolation.  Let us define the compression function $F_M$ as the
ratio of the final dark matter mass to the initial mass at the same
radius $xy(x)$:
\begin{equation}
  F_M(xy) \equiv {M_{\rm dm,f}(xy) \over M_{\rm dm,i}(xy)}
               = {M_{\rm dm,i}(x) \over M_{\rm dm,i}(xy)}.
  \label{eq:fm}
\end{equation}
We see that the increase of the dark matter mass due to baryonic
infall can be calculated from the initial dark matter profile $M_{\rm
dm,i}(x)$ given the transformation $y(x)$.  Since we can use an
analytic NFW profile, the accuracy of calculating $F_M(x)$ is similar
to that for $y(x)$.  Figure \ref{fig:fit_nfw} illustrates that
eq. (\ref{eq:fm}) provides an accurate fitting function at all radii
for all values of three independent parameters ($c$, $f_{\rm b}$,
$r_{\rm b}$).

\bigskip 

To summarize, in order to calculate the DM mass profile after 
contraction analytically one needs to: \\ 
(1) calculate the
maximum compression value $y_0$ using eqs.
(\ref{eq:y3sol}--\ref{eq:y0sol}); \\ (2) calculate the exponent $b$
of the weight function using eq. (\ref{eq:b}); \\ (3) calculate the
interpolating function $y(x)$ using eqs. (\ref{eq:t}--\ref{eq:ysol});
\\ (4) calculate the increase of the enclosed dark matter mass using
eq. (\ref{eq:fm}).

\bigskip

As mentioned above, the mass profiles of the NFW, exponential disk,
and Hernquist models are similar at $x \ll r_{\rm b},c^{-1}$: $m(x)
\propto x^2$.  This coincidence leads to a finite enhancement factor
for the dark matter mass and density in the central region: $M_{\rm
dm,f}(0)/M_{\rm dm,i}(0) = \rho_{\rm dm,f}(0)/\rho_{\rm dm,i}(0) =
y_0^{-2}$.  Therefore, the central contraction factor $y_0$
(eq. [\ref{eq:y0sol}]) provides a useful measure of the maximum
enhancement due to baryonic condensation.  The constant contraction
factor means that the inner slope of the dark matter distribution
after contraction is the same as before contraction. The radius at
which this slope is reached, however, depends on the halo
concentration and the baryon scale length (see Figure
\ref{fig:fit_nfw}).  At intermediate radii the post-contraction slope
is steeper. In general case, the slope of the final DM profile is
different from the initial profile and depends on the shape of the
baryon density profile.

\begin{figure}[t]
\centerline{\epsfysize3.5truein \epsffile{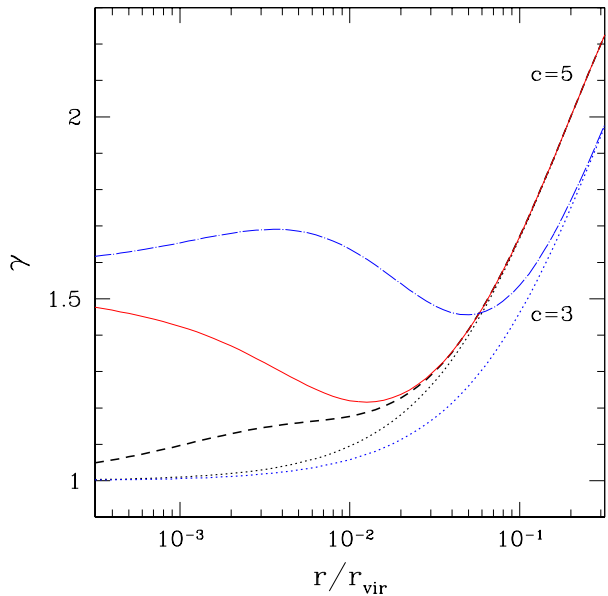}}
\caption{The slope of the dark matter profile in cluster MS 2137-23.
  Dotted lines are the initial NFW profile with $c=5$ and $c=3$.
  Dashed and solid lines are the post-contraction profile for the
  final baryon distributions given by the Hernquist model and Jaffe model,
  respectively, with $f_{\rm b} = 1.3\times 10^{-3}$ and $r_{\rm b}= 0.016$.
  Dot-dashed line shows
  the post-contraction profile in the case of the Jaffe model
  with $c=3$, $f_{\rm b} = 10^{-2}$ and $r_{\rm b}= 0.016$.
  \label{fig:slope_ms2137}}
\end{figure}

Since the value of the inner slope of the dark matter density profile
is important for testing CDM models against observations, it is
interesting to consider the change of the slope due to the
condensation of baryons into a configuration with an arbitrary final
density profile.  Let the baryon density at $x \ll 1$ be $\rho_{\rm
b}(x) \propto x^{-\nu}$ and therefore the mass $m_{\rm b}(x) \propto
x^{3-\nu}$.  If $\nu > 1$, the contraction would not be self-similar
with $x_f \propto x$ in the center.  Instead, the post-contraction
radius would scale as some power of the initial radius: $x_f \propto
x^{\alpha}$.  Substituting this into equation (\ref{eq:y}) and letting
$x \rightarrow 0$, we find $\alpha = (1+2w)/(1+(3-\nu)w)$.  The
post-contraction dark mass at the center would scale as $m(x) \propto
x^{2/\alpha}$, and therefore the density is
\begin{equation}
  \rho(x) \propto x^{-\gamma}, \quad \gamma = {1+2w\nu \over 1+2w}.
  \label{eq:gamma}
\end{equation}
For $\nu=1$, we recover our previous result, $\gamma=1$, which means
that after the contraction of NFW profile by baryons which form an
exponential or Hernquist profile, the asymptotic slope at small scales
remains the same.  For larger $\nu$, the slope $\gamma$ becomes
steeper than the initial slope.  Equation~(\ref{eq:gamma}) shows
(recall that $w\sim 1$) that for $\nu\sim 1-2$ the inner slope of the
dark matter profile will be quite close to the slope of the baryonic
profile.

\bigskip

As an illustration, we consider the profiles with the parameters
similar to those of the cluster MS 2137-23 described by
\citet{sand_etal02,sand_etal04}: the virial mass, radius, and
concentration are $M_{\rm vir} \sim 10^{15}\, \Msun$ and $r_{\rm vir}
\sim 2$ Mpc, and $c \approx 5$, respectively.  At the center of the
cluster lies a cD galaxy, which can be well fit by the Jaffe model:
$\rho_{\rm b} \propto x^{-2} (x+r_{\rm b})^{-2}$ with $r_{\rm b}
\approx 33$ kpc.  Assuming the best fitting mass-to-light ratio $M/L_V
= 2.6$ \citep{sand_etal04}, the mass of the central galaxy is $1.2
\times 10^{12}\, \Msun$.  In our notation, the relevant parameters are
$f_{\rm b} = 1.2\times 10^{-3}$, $r_{\rm b} = 0.016$.

Figure \ref{fig:slope_ms2137} shows the slope of the post-contraction
dark matter density profile appropriate for MS 2137-23 in the range
$10^{-3} < x < 10^{-1}$ which is used in the Sand et al. analysis.  If
the baryon distribution was described by a Hernquist model ($\nu=1$),
the inner slope would tend to $\gamma=1$, as we discussed above.  For
the Jaffe model ($\nu=2$), on the other hand, the slope {\it steepens}
at $x \lesssim 10^{-2}$.  According to equation (\ref{eq:gamma}), the
asymptotic inner slope is $\gamma \approx 1.6$.  This slope is reached
faster if we consider a more massive galaxy ($f_{\rm b} = 10^{-2}$)
and a less concentrated cluster ($c=3$).

\twocolumn
\bibliography{contra}

\end{document}